\documentclass[aps,preprintnumbers,amsmath,amssymb,prl,superscriptaddress,longbibliography,twocolumn]{revtex4-1}
\usepackage{dcolumn}% Align table columns on decimal point
\usepackage{color}
\usepackage{caption}
\usepackage[caption=false]{subfig}
\usepackage{feynmp}
\usepackage{feynmp-auto}
\usepackage{systeme,mathtools}
\usepackage{stmaryrd}
\usepackage{easyReview}

\def\comment#1{}

\newcommand{\nc}{\newcommand}
\nc{\beq}{\begin{eqnarray}}
	\nc{\eeq}{\end{eqnarray}}
\nc{\scs}{\scriptstyle}
\nc{\setval}{\fmfset{wiggly_len}{3mm} \fmfset{arrow_len}{1.5mm}
	\fmfset{arrow_ang}{13} \fmfset{dash_len}{1.5mm}\fmfpen{0.125mm}
	\fmfset{dot_size}{2thick}}

\usepackage{bm,latexsym,mathrsfs,enumerate,color}
\usepackage[mathcal]{euscript}
\usepackage[breaklinks=true,unicode=true,urlcolor = blue,colorlinks = true,citecolor = blue,linkcolor = blue]{hyperref}
\usepackage{graphicx}
\usepackage{todonotes}
\usepackage{wrapfig}
\usepackage{easyReview}

\renewcommand{\vec}[1]{\bm{#1}}

\def\slashchar#1{\setbox0=\hbox{$#1$}           % set a box for #1
	\dimen0=\wd0                                 % and get its size
	\setbox1=\hbox{/} \dimen1=\wd1               % get size of /
	\ifdim\dimen0>\dimen1                        % #1 is bigger
	\rlap{\hbox to \dimen0{\hfil/\hfil}}      % so center / in box
	#1                                        % and print #1
	\else                                        % / is bigger
	\rlap{\hbox to \dimen1{\hfil$#1$\hfil}}   % so center #1
	/                                         % and print /
	\fi}                                         %

\DeclareMathAlphabet\mathbfcal{OMS}{cmsy}{b}{n}

\def\sigmab{{\mbox{\boldmath $\sigma$}}}

\begin{document}
	
	\title{Frozen deconfined quantum criticality}
	
	\author{Vira Shyta}
	\affiliation{Institute for Theoretical Solid State Physics, IFW Dresden, Helmholtzstr. 20, 01069 Dresden, Germany}
	\affiliation{KAU — Department of Theoretical and Mathematical Physics, Kyiv Academic University, 36 Vernadsky blvd., Kyiv 03142, Ukraine}
	
	\author{Jeroen van den Brink}
	\affiliation{Institute for Theoretical Solid State Physics, IFW Dresden, Helmholtzstr. 20, 01069 Dresden, Germany}
	\affiliation{Institute for Theoretical Physics and W\"urzburg-Dresden Cluster of Excellence ct.qmat, TU Dresden, 01069 Dresden, Germany}
	
	\author{Flavio S. Nogueira}
	\affiliation{Institute for Theoretical Solid State Physics, IFW Dresden, Helmholtzstr. 20, 01069 Dresden, Germany}
	
	\begin{abstract}
		There is a number of contradictory findings with regard to whether the theory describing easy-plane quantum antiferromagnets undergoes a second-order phase transition. The traditional Landau-Ginzburg-Wilson approach suggests a first-order phase transition, as there are two different competing order parameters. On the other hand, it is known that the theory has the property of self-duality which has been connected to the existence of a deconfined quantum critical point (DQCP).  The latter regime suggests that order parameters are not the elementary building blocks of the theory, but rather consist of fractionalized particles that are confined in both phases of the transition and only appear — deconfine — at the critical point.  Nevertheless, many numerical Monte Carlo simulations disagree with the claim of a DQCP in the system, indicating instead a first-order phase transition.  
		Here we establish from exact lattice duality transformations and renormalization group analysis that the easy-plane CP$^1$ antiferromagnet does feature a DQCP. We uncover the criticality starting from a regime analogous to the zero temperature limit of a certain classical statistical mechanics system which we therefore dub  “frozen”. At criticality our bosonic theory is dual to a fermionic one with two massless Dirac fermions, which thus undergoes a second-order phase transition as well.
		
	\end{abstract}
	
	\maketitle
	
	Quantum antiferromagnets that possess a global $SU(2)$ symmetry and an 
	emergent $U(1)$ gauge 
	symmetry can give rise to exotic phases of matter, like spin liquids and 
	valence-bond solid states \cite{fradkin2013field,Sachdev-book}. An interesting scenario 
	occurs when at low temperatures the system features a quantum critical point at a  
	value $g_c$ of some effective coupling constant. For instance, such a quantum 
	critical point can separate the magnetically ordered (N\'eel) state, which 
	breaks the $SU(2)$ symmetry, from a dimerized paramagnetic state breaking the lattice symmetries. 
	The latter finds a paradigmatic realization in the valence-bond solid 
	(VBS) phase 
	\cite{Read-Sachdev_PhysRevB.42.4568,Sachdev-book}. An effective field theory formulation 
	with an emergent $U(1)$ gauge symmetry is achieved in this context by rewriting 
	the unit vector field $\vec{n}$ representing the direction of the magnetization in 
	terms of a doublet of complex fields, $z_a$ ($a=1,2$), $\vec{n}=z_a^*\sigmab_{ab}z_b$, where $|z_1|^2+|z_2|^2=1$ and $\sigmab$ is a vector of Pauli matrices and summation over repeated indices is implied.  A global $O(3)$ symmetry becomes henceforth a global $O(4)$ one under this mapping. It also makes the $U(1)$ symmetry manifest, since $\vec{n}$ is invariant under 
	the local gauge transformation $z_a(x)\to e^{i\theta(x)}z_a(x)$.
	This map, which is also referred to as a CP$^1$ representation, leads to a lattice gauge theory of quantum antiferromagnets. Under some very precise circumstances 
	the magnetization falls apart in such a theory, liberating more elementary modes — spinon fields $z_a$. This regime leads to a special type of universal behavior, governed by the so called deconfined 
	quantum criticality (DQC) \cite{Senthil1490,Senthil_PhysRevB.70.144407}. 
	
	A salient property of the DQC paradigm is that it describes a universality class 
	that cannot be derived from a Landau-Ginzburg-Wilson type of approach.  
	In the latter a transition between phases breaking different symmetries (competing 
	orders), be it at zero or finite temperature, is a first-order one, which implies that the theory should not feature 
	a critical point. 
	Within the DQC theory, on the other hand, a second-order phase 
	transition is predicted to occur.  
	Furthermore, DQC predicts a large anomalous 
	dimension $\eta_N$ for the correlator $G(x)=\langle\vec{n}(x)\cdot\vec{n}(0)\rangle$ at 
	the critical point \cite{Senthil1490,Senthil_PhysRevB.70.144407}, this result being a 
	consequence of the composite field character of $\vec{n}$ underlying the CP$^1$ representation.  
	This prediction has been confirmed by multiple computer simulations on a number of specific lattice models 
	proposed to describe DQC  \cite{Sandvik_PhysRevLett.104.177201,Melko-Kaul_PhysRevLett.100.017203,Nahum_PhysRevX.5.041048,Sandvik_PhysRevX.7.031052}. 
	
	A paradigmatic model argued to exhibit DQC features an easy-plane anisotropy that reduces
	the global $O(4)$ symmetry to an Abelian one, namely, $U(1)\times U(1)$ \cite{Senthil1490,Senthil_PhysRevB.70.144407}. The question of whether the anisotropic quantum antiferromagnets feature a deconfined critical point has been open since the creation of the field \cite{HLM_PhysRevLett.32.292,Hikami,Herbut-Tesanovic_PhysRevLett.76.4588,  KUKLOV20061602,Kragset_PhysRevLett.97.247201, Nogueira_PhysRevB.76.220403, Sudbo_PhysRevB.82.134511, Pollmann_PhysRevLett.120.115702,Sandvik_PhysRevX.7.031052,  Senthil_PhysRevX.7.031051, Kaul_Desai, Kaul_ep_FP, Shyta_PhysRevLett.127.045701}.
	This model has the advantage of being analytically tractable to a certain extent and  has been demonstrated to exhibit a
	self-dual regime \cite{Motrunich-Vishwanath_PhysRevB.70.075104}. 	 Interestingly, from the symmetry point of view, this model can also describe the phase transition in two-component superconductors 
	\cite{Sudbo_PhysRevB.82.134511}. Early computer simulations 
	\cite{KUKLOV20061602,Kragset_PhysRevLett.97.247201} have failed 
	to find evidence of a second-order phase transition in this case. 
	Nevertheless, it has been recently suggested in the context of bosonization dualities 
	that  DQC can be achieved in this self-dual model \cite{Senthil_PhysRevX.7.031051}. 
	Also, more recently, other numerical works \cite{Pollmann_PhysRevLett.120.115702,Sandvik_PhysRevX.7.031052} on easy-plane systems concluded that a second-order phase transition occurs. However, the controversy persists, as a recent numerical work \cite{Kaul_Desai} for the easy-plane $J-Q$ model favors a first-order phase transition.   	 
	
	Here we demonstrate by purely analytical means that the easy-plane model in CP$^1$ representation features DQC in quite a specific regime. The key observation is that by considering the lattice theory  as a classical statistical mechanics model, we identify the coupling constant $g$ as playing the role of temperature in the action. 
From considering the “frozen” $g\to 0$ regime, where we show  that a quantum critical point exists, we construct a duality that allows us to derive DQC in the more general case ($g\neq 0$). The standard duality transformation in the spirit of Refs. \cite{Peskin1978,Dasgupta-Halperin_PhysRevLett.47.1556,THOMAS1978513} performed in the frozen regime leads to a $U(1)\times U(1)$-symmetric Higgs theory featuring two gauge fields 
%coupled by a flux attachment term, 
and no Maxwell terms. To access the $g\neq 0$ case, however, we dualize only a single $U(1)$ sector, obtaining the same model but with Maxwell terms and different gauge couplings. A subsequent renormalization group (RG) analysis then establishes the existence of a quantum critical point. Importantly, the obtained critical regime is the same as in the frozen limit. We demonstrate that at criticality the theory is topologically ordered and is dual to a theory of two massless Dirac fermions in the infrared (IR) limit. From the derived bosonization duality we conclude that also the fermionic theory possesses a quantum critical regime. 
		 
	The natural starting point to investigate DQC lies in quantum antiferromagnetic systems, whose behavior can be described in terms of two spinon fields $z_I$, $I=1,2$, via the CP$^1$ representation of the nonlinear $\sigma$ model \cite{Kuklov2008},
	\begin{equation}
		S = - \frac{1}{g} \sum_{\langle ij\rangle, I} \left( z^*_{Ii}z_{Ij}e^{i a_{ij}}+ c.c.\right) +\frac{1}{2e^2}\sum_{j}(\epsilon_{\mu\nu\lambda}\Delta_\nu a_{j\lambda})^2 ,
	\end{equation}
	with a local constraint $|z_{1j}|^2 + |z_{2j}|^2 =1$. The first term in $S$ describes nearest-neighbor hopping of the CP$^1$ fields on a square lattice and  
	$a_{ij}$ is an emergent gauge field. The second term represents a Maxwell Lagrangian in the lattice.  The so called  deep easy-plane limit  fixes the spinon amplitudes to be equal, which leads to $|z_1|^2 = |z_2|^2 =1/2$ due to the initial constraint. In this way the action takes the form,
	\begin{equation}
		\label{Eq:XY-gauged}
		S = - \frac{1}{2g} \sum_{\langle ij\rangle, I}  \cos{(\theta_{Ii}-\theta_{Ij} - a_{ij})}  +\frac{1}{2e^2}\sum_{j}(\epsilon_{\mu\nu\lambda}\Delta_\nu a_{j\lambda})^2 ,
	\end{equation}
	where $\theta_{Ii}$ arise from the polar representation of the spinons $z_{Ij} = \rho_I e^{i \theta_{Ij}}$ with $\rho_I=1/\sqrt{2}$. We recognize the theory as a gauged version of a two-component $XY$ model. When interpreting this Euclidean action as a classical statistical physics Hamiltonian, the coupling constant $g$ plays a role analogous to the temperature.
	The action (\ref{Eq:XY-gauged}) can be well approximated in the form of a $U(1)\times U(1)$ lattice Villain system  
	\cite{Motrunich-Vishwanath_PhysRevB.70.075104}, 
	\begin{eqnarray}
		\label{Eq:Action}
		S&=&\frac{1}{2}\sum_j\left[
		\frac{1}{g}\sum_{I=1,2}(\Delta_\mu\theta_{Ij}-2\pi n_{Ij\mu}-a_{j\mu})^2
		\right.\nonumber\\
		&+&\left.\frac{1}{e^2}(\epsilon_{\mu\nu\lambda}\Delta_\nu a_{j\lambda})^2\right],
	\end{eqnarray}
	where $\Delta_\mu$ is the lattice derivative, 
	$n_{Ij\mu}$ are integer-valued lattice fields and
	$\theta_{Ij}\in[-\pi,\pi]$. The action (\ref{Eq:Action}) has besides the usual gauge invariance $\theta_{Ij}\to\theta_{Ij}+\alpha_j$,  $a_{j\mu}\to a_{j\mu}+\Delta_\mu\alpha_j$, two $\mathbb{Z}$ gauge symmetries, $n_{Ij\mu}\to n_{Ij\mu}+\Delta_\mu K_{Ij}$, $\theta_{Ij}\to\theta_{Ij}+2\pi K_{Ij}$, for integers $K_{Ij}$. 

	In order to obtain the dual theory, we use the Poisson summation formula \cite{Poisson-formula,Peskin1978} to introduce a new integer-valued field, 
	\begin{eqnarray}
		\label{eq:Poisson}
		&&\sum_{\{n_{Ij\mu}\}}e^{-\frac{1}{2g}(\Delta_\mu\theta_{Ij}-2\pi n_{Ij\mu}-a_{j\mu})^2}
		\nonumber\\
		&\sim&\sum_{\{N_{Ij\mu}\}}e^{\frac{g}{2}N_{Ij\mu}^2+iN_{Ij\mu}(\Delta_\mu\theta_{Ij}
			-a_{j\mu})},
	\end{eqnarray}
	which allows us to integrate out $\theta_{Ij}$ to obtain the constraints 
	$\Delta_\mu N_{Ij\mu}=0$. Thus, after the constraints are solved by 
	$N_{Ij\mu}=\epsilon_{\mu\nu\lambda}\Delta_\nu M_{Ij\lambda}$ and $a_{j\mu}$ is 
	integrated out, we obtain the dual action in the form \cite{Senthil_PhysRevB.70.144407,Smiseth_PhysRevB.71.214509},  
	\begin{eqnarray}
		\label{Eq:dual-action-dqc}
		\widetilde{S}&=&\sum_j\left\{\sum_{I=1,2}\left[\frac{g}{2}(\epsilon_{\mu\nu\lambda}
		\Delta_\nu b_{Ij\lambda})^2-2\pi im_{Ij\mu}b_{Ij\mu}\right]
		\right.\nonumber\\
		&+&\left.\frac{e^2}{2}(b_{1j\mu}+b_{2j\mu})^2
		\right\},
	\end{eqnarray}
	where we have used the Poisson formula once more to promote the integer-valued fields 
	$M_{Ij\mu}$ to real-valued fields $b_{Ij\mu}$ and 
	the constraints $\Delta_\mu m_{Ij\mu}=0$ hold. 
	Physically the fields $m_{Ij\mu}$ represent vortices and the zero divergence 
	constraints imply that all vortex lines form loops \cite{kleinert1989gauge}.  

	We will consider now the ``zero temperature” limit ($g\to 0$) of the obtained dual model. This causes the Maxwell terms to vanish. After integrating out the gauge field $b_{2i\mu}$  and solving the constraint $\Delta_\mu m_{2j\mu} = 0$ via the integral representation of the Kronecker  $\delta$, the following action is obtained through the Poisson summation, 
\begin{eqnarray}
	\label{Eq:dual-action-dqc-frozen-1gauge-1}
	\widetilde{S}&=&\sum_j\left\{\frac{e^2}{8 \pi^2}\left(\Delta_\mu\widetilde{\theta}_j - 2 \pi \widetilde{n}_{j\mu}- 2 \pi b_{1j\mu}\right)^2\right.
	\nonumber\\
	&-& \left.2\pi i m_{1j\mu}b_{1j\mu}
	\right\},
\end{eqnarray}
 where the Poisson summation formula was applied to promote the integer-valued field to be real-valued. 
 
 By performing a shift (``Higgsing") $b_{1j\mu}\to b_{1j\mu}+(\Delta_\mu\widetilde{\theta}_j - 2 \pi \widetilde{n}_{j\mu})/(2\pi)$, the action of Eq. (\ref{Eq:dual-action-dqc-frozen-1gauge-1}) becomes simply $\widetilde{S}=\sum_j\left(e^2b_{1j\mu}^2/2-2\pi im_{1j\mu}b_{1j\mu}\right)$, since the zero divergence constraint on $m_{1j\mu}$ makes $\widetilde{\theta}_j$ disappear and the term $2\pi im_{1j\mu}\widetilde{n}_{j\mu}$ does not contribute as its exponential yields the unity. After integrating $b_{1j\mu}$ out a vortex loop gas representation of the $XY$ model is obtained in a way akin to the one considered in Refs. \cite{Peskin1978,Dasgupta-Halperin_PhysRevLett.47.1556}. In this case $e^2/(2\pi)^2$ plays the role of the inverse temperature. Therefore, the frozen regime $g\to 0$ has a quantum critical point in the inverted $XY$ universality class \cite{Dasgupta-Halperin_PhysRevLett.47.1556}. 
 
  Let us now show that the quantum critical regime associated to the frozen limit exhibits topological order. This is in contrast to the standard inverted $XY$ universality class, where no such an order arises. To demonstrate the topological order underlying the action of Eq. (\ref{Eq:dual-action-dqc-frozen-1gauge-1}), we first solve  the zero divergence constraint on $m_{1j\mu}$ via $m_{1j\mu} = \epsilon_{\mu\nu\lambda}\Delta_\nu \widetilde{N}_{j\lambda}$, and promote the field  $ \widetilde{N}_{j\mu}$ to be real-valued via the Poisson formula, to obtain, 
 	\begin{eqnarray}
 		\label{Eq:dual-action-dqc-frozen-1gauge-2}
 		\widetilde{S'}&=&\sum_j\left\{\frac{e^2}{8 \pi^2}\left(\Delta_\mu\widetilde{\theta}_j - 2 \pi \widetilde{n}_{j\mu}- 2 \pi b_{1j\mu}\right)^2\right.\nonumber\\&-&\left. 2\pi i b_{1j\mu} \epsilon_{\mu\nu\lambda}\Delta_\nu h_{j\lambda}- 2\pi i \widetilde{m}_{j\mu} h_{j\mu}
 		\right\}.
 	\end{eqnarray}
 	The field equation for $b_{1j\mu}$ expresses the fact that the Noether current $J_{j\mu}=(e^2/4\pi^2)(\Delta_\mu\widetilde{\theta}_j - 2 \pi \widetilde{n}_{j\mu}- 2 \pi b_{1j\mu})$ is topological, since $J_{j\mu}=2\pi i \epsilon_{\mu\nu\lambda}\Delta_\nu h_{j\lambda}$. 
 	
 	We now add a term  $\varepsilon\widetilde{m}^2_{j\mu}/2$, where $\varepsilon$ represents  the vortex core energy \cite{kleinert1989gauge}, which can also be viewed as a chemical potential for the vortex loops \cite{Peskin1978,Dasgupta-Halperin_PhysRevLett.47.1556,THOMAS1978513, Jose_PhysRevB.16.1217}. Introducing a phase field $\widetilde{\varphi}_j$ from the integral representation of the Kronecker delta constraint on $\widetilde{m}_{j\mu}$, we arrive at the following action,
 	\begin{eqnarray}
 		\label{Eq:dual-action-dqc-frozen-2componentHiggs}
 		\widetilde{S'}&=&\sum_j\left\{\frac{e^2}{8 \pi^2}\left(\Delta_\mu\widetilde{\theta}_j -  2 \pi\widetilde{n}_{1j\mu}- b_{1j\mu}\right)^2\right.\nonumber\\&+&\left.\frac{1}{2\varepsilon}\left(\Delta_\mu\widetilde{\varphi}_j - 2 \pi \widetilde{n}_{2j\mu}- b_{2j\mu}\right)^2\right.\nonumber\\&-&\left. \frac{i}{2\pi} b_{1j\mu} \epsilon_{\mu\nu\lambda}\Delta_\nu b_{2j\lambda}
 		\right\},
 	\end{eqnarray}
 	where a rescaling, $b_{Ij\mu}\to b_{Ij\mu}/(2\pi)$ has been made. 
 	This formulation of the dual action in the frozen regime has a number of interesting features. First, we note that thanks to the so called BF term [the last term in Eq. (\ref{Eq:dual-action-dqc-frozen-2componentHiggs})], both Noether currents are topological in view of the field equations for both $b_{1j\mu}$ and $b_{2j\mu}$. Second and also in view of the property just mentioned, the dual action (\ref{Eq:dual-action-dqc-frozen-2componentHiggs}) can be regarded as a theory  for topologically ordered superconductors in 2+1 dimensions \cite{Hansson-TopOrderSCs}. In this interpretation, one of the Noether currents is associated to the quasi-particle currents while the other one describes the vortex current. We therefore conclude that such a topologically ordered system undergoes a second-order phase transition governed by the inverted $XY$ universality class. 

 	So far, using the exact duality transformations in the frozen limit, we showed the existence of an XY critical point and demonstrated its topological nature. Grounded in these findings, we will now expand the criticality claim to the case where the coupling $g$ is finite.  
	 	 In order to do so, we develop a new strategy where only one $U(1)$ sector of the easy-plane CP$^1$ model is dualized. This approach is motivated by  the intuition we developed  considering the frozen dual model. Indeed, we note that the frozen limit causes the dual model to have one vortex loop field suppressed, as is seen from Eq. \eqref{Eq:dual-action-dqc-frozen-1gauge-1}.
	 The procedure will allow us to demonstrate the existence of a quantum critical point starting from a finite $g$. 
	
	Returning to the easy-plane CP$^1$ model of Eq. \eqref{Eq:Action},  we repeat the step discussed in Eq. \eqref{eq:Poisson} but only for one phase variable  (we choose $\theta_{2j}$). This leads to
	 \begin{eqnarray}
		\label{Eq:semidual-action-dqc-0}
		\widetilde{S}''&=&\sum_j\left[\frac{1}{2g}\left(\Delta_\mu\theta_{1j} - 2 \pi n_{1j\mu}-  a_{j\mu}\right)^2\right.\nonumber\\&+&\left.\frac{1}{2e^2}\left(\epsilon_{\mu\nu\lambda}\Delta_\nu a_{j\lambda}\right)^2+i a_{j\mu}\epsilon_{\mu\nu\lambda}\Delta_\nu b_{j\lambda} \right.\nonumber\\&+&\left. \frac{g}{2}\left(\epsilon_{\mu\nu\lambda}\Delta_\nu b_{j\lambda}\right)^2 - 2 \pi i b_{j\mu} m_{j\mu}
		\right],
	\end{eqnarray}
where $b_{j\mu}$ is a new gauge field and $m_{j\mu}$ is a lattice vortex loop field. 

Similarly to our previous calculations,  the constraint $\Delta_{\mu}m_{j\mu}=0$ allows us to introduce a new phase field $\varphi_j$. Adding the vortex core energy and using the Poisson summation formula, we arrive at the following action,
\begin{eqnarray}
	\label{Eq:semidual-action-dqc-1}
		\widetilde{S}''&=&\sum_{j}\left[\frac{1}{2 g}\left(\Delta_{\mu} \theta_{1j}-2 \pi n_{j\mu}-a_{j \mu}\right)^{2}\right.\nonumber\\
		&+&\left.\frac{1}{2 \varepsilon}\left(\Delta_{\mu} \varphi_{j}-2 \pi \tilde{n}_{j \mu}-b_{j \mu}\right)^{2} \right.\nonumber\\
		&+&\left.\frac{g}{8 \pi^{2}}\left(\varepsilon_{\mu \nu \lambda} \Delta_{\nu} b_{j \lambda}\right)^{2}+\frac{1}{2 e^{2}}\left(\varepsilon_{\mu \nu \lambda} \Delta_{\nu} a_{j \lambda}\right)^{2}\nonumber\right.\\&+&\left.\frac{i}{2 \pi} \varepsilon_{\mu \nu \lambda} a_{j \mu} \Delta_{\nu} b_{j \lambda}\right],
\end{eqnarray}
where the new gauge field was rescaled, $b_{j\mu}\to b_{j\mu}/(2\pi)$. The model above is reminiscent of the one  obtained in the frozen regime in Eq. \eqref{Eq:dual-action-dqc-frozen-2componentHiggs} as both actions contain two Higgs terms and two  gauge fields coupled via a topological BF term. The crucial difference, however, lies in the fact that  in Eq. \eqref{Eq:semidual-action-dqc-1} the Maxwell terms for both $a_{j\mu}$ and $b_{j\mu}$ are present.  Hence, these models actually represent different physical pictures, as we will discuss in more detail below.

Let us first put into perspective the physical significance of the action (\ref{Eq:semidual-action-dqc-1}) and recapitulate what we have achieved so far. 
We started with the $U(1)\times U(1)$ gauge theory of Eq. (\ref{Eq:Action}) (or any of its equivalent forms), and derive the exact dual action seen in Eq.   (\ref{Eq:dual-action-dqc}) which features two (dual) gauge fields. 
Then we show that in the frozen limit the dual action can be cast in the form (\ref{Eq:dual-action-dqc-frozen-2componentHiggs}) with the gauge fields coupled via a BF term. This theory describes an ensemble of two types of vortex loops \textit{having the same gauge charge}, as the particle-vortex duality has been performed in \textit{both} $U(1)$ sectors. By contrast, Eq. (\ref{Eq:semidual-action-dqc-1}) results from performing the particle-vortex duality in only one $U(1)$ sector. This naturally implies that the gauge charge of the particles in one $U(1)$ sector is attached to the flux resulting from particle-vortex duality in the other $U(1)$ sector. Indeed, now the gauge coupling of the dualized $U(1)$ sector corresponds to the phase stiffness of the original particles, while the $U(1)$ sector that has not been dualized still retains its original ``electric" charge $e$. In this sense, the action of Eq. (\ref{Eq:semidual-action-dqc-1}) represents rather an electric-magnetic duality in 2+1 dimensions. Note that due to the presence of Maxwell terms, Noether currents are no longer topological, in contrast to the frozen regime. This causes the gauge potentials to be gapped, similarly to the situation of (2+1)-dimensional superconductors where the topological (BF) action has to be supplemented with Maxwell terms in order to account for the plasmon modes \cite{Hansson-TopOrderSCs}. Here the mutually dual Maxwell terms appear quite naturally as a consequence of the duality transformation.

From Eq. \eqref{Eq:semidual-action-dqc-1} we infer the continuum field theory Lagrangian in imaginary time,
	\begin{eqnarray}
		\label{Eq:L-semidual}
		\widetilde{\mathcal{L}}''&=&
		|(\partial_\mu-iea_{\mu})\phi_1|^2+m^2|\phi_1|^2+ \frac{u_1}{2}|\phi_1|^4 \nonumber\\
		&+&|(\partial_\mu-i\frac{2 \pi}{\sqrt{g}}b_{\mu})\phi_2|^2+m^2|\phi_2|^2+ \frac{u_2}{2}|\phi_2|^4 
		\nonumber\\
		&+&\frac{1}{2}\left(\varepsilon_{\mu \nu \lambda} \partial_{\nu} a_{ \lambda}\right)^{2}+ \frac{1}{2}\left(\varepsilon_{\mu \nu \lambda} \partial_{\nu} b_{ \lambda}\right)^{2}\nonumber\\&+&\frac{i e}{\sqrt{g}} a_{\mu} \epsilon_{\mu\nu\lambda}\partial_\nu b_{\lambda},
	\end{eqnarray}
where the gauge fields were rescaled as $a_\mu \to e a_\mu$ and $b_{\mu}\to \frac{2 \pi}{\sqrt{g}}b_{\mu}$. This two-component Lagrangian features two different charges: $e$ from the original model and $\widetilde{e} = \frac{2 \pi}{\sqrt{g}}$ obtained for the dual $U(1)$ sector. 

The RG analysis performed on the Lagrangian \eqref{Eq:L-semidual} yields the interesting result 
%that the critical theory belongs to the same universality class as the frozen model with two dual $U(1)$ sectors
that the theory features a critical regime which belongs to the same universality class as the frozen model with two dual $U(1)$ sectors
(full details of calculations can be found in the Supplemental Material (SM) \cite{SM}).
We define the renormalized dimensionless couplings of the $|\phi|^4$-interactions  as $\hat{u}_I=u_{IR}m_R^{-\epsilon}$ for $I=1, 2$, 
where $m_R$ is the renormalized mass. Here we have also introduced $\epsilon=4-d$ for a spacetime dimension $2<d\leq 4$ in order to obtain a perturbative fixed point of $\mathcal{O}(\epsilon)$.  The one-loop $\beta$ functions for both $\hat{u}_I$ have the same form and are given by $\beta_{{\hat{u}}_I}=-\epsilon\hat{u}_I+5 \hat{u}_I^2/(8 \pi)$. 
 The  IR stable fixed points, $\hat{u}^*_I = (8 \pi)\epsilon/5$, obtained from the RG equations are consistent with the $XY$ universality class. 
 Let us mention that an RG analysis of the Dasgupta-Halperin dual model also implies an XY fixed point \cite{Herbut_1997}. In both cases, this occurs due to a gapped gauge field. However, the mechanism by which the gauge fields of Eq. \eqref{Eq:semidual-action-dqc-1}  become massive is quite different from the one described in Ref. \cite{Herbut_1997}. In fact, the theory above is gauge invariant and the gap follows from the presence of a topological BF term. 

 The critical point is reached when both dimensionless counterparts of the renormalized couplings $e_R^2$ and $\widetilde{e}^2_R$ flow to their fixed points as well as  $\hat{u}_I \to (8 \pi)\epsilon/5$. The  $\beta$ functions for the gauge couplings  calculated from the one-loop vacuum polarization have the same general form, $\beta_{f} = - \epsilon f + f^2/(24 \pi)$, where $f^2$ is a dimensionless renormalized coupling corresponding to either $f = e_R^2m_R^{-\epsilon}$ or $f =\widetilde{e}^2_R m_R^{-\epsilon}$. From the $\beta$ functions it is straightforward to find the IR stable fixed points, $\hat{e}^2_* = 24 \pi \epsilon $ and $1/\hat{g}_* = \epsilon \pi/6$, where   $\hat{e}^2$ and $\hat{g}$ are dimensionless couplings. When the $\beta$ functions vanish, the RG flows of $\hat{e}^2$ and $\hat{g}$ are dual with respect to each other, $\beta_{\hat{e}^2}/\hat{e}^2 = - \beta_{\hat{g}}/\hat{g}$. This leads to a Dirac-like relation,  $\hat{e}^2 \hat{g} = 4 \pi^2$, which is satisfied at the fixed point.  Importantly, from this analysis it follows that at criticality the Maxwell terms in the Lagrangian \eqref{Eq:L-semidual} become RG irrelevant. Consequently, we conclude that this critical theory belongs to the same universality class as a continuous version of the frozen dual model in Eq. \eqref{Eq:dual-action-dqc-frozen-2componentHiggs} {\it where both $U(1)$ sectors are dualized}. Hence, a continuum field theory implied by Eq. \eqref{Eq:dual-action-dqc-frozen-2componentHiggs} can be readily identified to Eq. (\ref{Eq:L-semidual}) with the Maxwell terms absent and $\hat{e}^2 \hat{g} = 4 \pi^2$. Incidentally, since an RG analysis in terms of {\it bare} rather than renormalized parameters leads to the same critical behavior \cite{zinn2021quantum}, we obtain that a dimensionless bare coupling defined by $\hat{g}_0=g\Lambda$ causes $g$ to flow to zero as the fixed point $g_0^*\neq 0$ is approached when the ultraviolet cutoff $\Lambda\to\infty$. Although this is the same fixed point we have obtained for the dimensionless renormalized coupling (as implied by scale invariance), the result that $g\to 0$ as $\Lambda\to\infty$ highlights the role played by the frozen regime.       

From the one-loop RG analysis it follows that the correlation length critical exponent $\nu\approx 0.625$ (after setting $\epsilon=1$), which is precisely the  one-loop value for the $XY$ universality class.  Furthermore, the salient critical property of DQC is the large anomalous scaling dimension of order parameters. From the irrelevance of the Maxwell terms near the IR fixed points, we see that the correlation function of the VBS order parameter at the quantum critical point can be represented as a bound state between a vortex and a particle (here a spinon) operator. The correlation function   
associated to the VBS order parameter is  the gauge invariant correlation 
function $C_{\rm VBS}(x)=\langle\phi_1^*(x)\phi_2(x)\phi_2(0) \phi_1^*(0)\rangle$.
The anomalous dimension $\widetilde{\eta}$ is defined via the large distance behavior $C(x)\sim 1/|x|^{1+\widetilde{\eta}}$  
at the critical point.  We obtain that $\widetilde{\eta}=1-2\eta_{12}$, where  $\eta_{12}$ is the anomalous dimension of the gauge-invariant operator $\phi_1^*(x)\phi_2(x)$ \cite{zinn2021quantum}. 
At one-loop order we obtain $\eta_{12}=-\hat{u}_*/(8\pi)$ and, therefore,  $\widetilde{\eta}^{U(1)}=1.4$. The result shows that we are dealing with a modified $XY$ universality class,  akin to the so called $XY_*$  discussed in Ref. \cite{Isakov193}, where  an anomalous dimension $\eta=1.493$ is numerically obtained for a lattice  boson model exhibiting fractionalized excitations.

	So far we have demonstrated that actions describing the frozen DQC naturally contain a topological BF term  linked  to a topological order arising at the critical point. We will now explore the interesting fact that such a BF term flux attachment allows one to derive a duality within a bosonization framework \cite{Karch_PhysRevX.6.031043, Nahum_PhysRevX.5.041048, Senthil_PhysRevX.7.031051}. Using this technique, we show in the SM \cite{SM} that the bosonic two-component model with dynamical gauge fields coupled via a BF term is dual to the theory of two massless Dirac fermions coupled via a shared gauge field,
\begin{eqnarray}
	\label{Eq:MainBosonization}
	\mathcal{L}_b&=&\sum_{I=1,2}\left[|(\partial_\mu-ib_{I\mu})\phi_I|^2 + m^2 |\phi_I|^2 + \frac{u}{2}|\phi_I|^4 \right]\nonumber\\&+& \frac{i}{2\pi}\epsilon_{\mu\nu\lambda}b_{1\mu}\partial_\nu b_{2\lambda}
	\nonumber\\
	&&\hspace{2cm} \big\Updownarrow
	\nonumber\\
	\mathcal{L}_f&=&\sum_{I=1,2} \bar{\psi}_I(\slashchar{\partial}-i\slashchar{a})\psi_I.
\end{eqnarray}
Thus, the duality integrates the topologically ordered $U(1)\times U(1)$ Abelian Higgs model into a wider duality web. As the bosonization duality leads to the expectation that critical behavior on both sides is the same,  we conclude that the fermionic side of the duality also undergoes a second-order phase transition. If we now are to consider the fermionic theory as an intermediate step, we obtain a boson-boson duality between the easy-plane CP$^1$ model and its dual version capturing DQC. 
	
 In summary we have analyzed the DQC paradigm for 
	the easy-plane antiferromagnet by exploring the interplay between duality transformations and the RG scaling behavior. We have identified a quantum critical regime  given 
	by a modified $XY$ universality class, where at the fixed point $\hat{e}^2\hat{g}=(2\pi)^2$. Furthermore, at the critical point the topological order which arises in the frozen regime is recovered.

	%\subsection*{Single column equations}
	
	%Authors may use 1- or 2-column equations in their article, according to their preference.
	
	%To allow an equation to span both columns, use the \verb|\begin{figure*}...\end{figure*}| environment mentioned above for figures.
	
	%Note that the use of the \verb|widetext| environment for equations is not recommended, and should not be used. 

	\begin{acknowledgments}
		We thank the Deutsche Forschungsgemeinschaft (DFG) 
		for support through the W\"urzburg-Dresden Cluster of Excellence on Complexity and Topology in Quantum Matter – ct.qmat (EXC 2147, project-id 39085490) and 
		the Collaborative Research Center  
		SFB 1143 (project-id 247310070). V.S. has been supported by UKRATOP-project (funded by BMBF with  grant number 01DK18002). 
	\end{acknowledgments}  
	
	\bibliography{frozen-dqc-refs}
	
	\section{Supplemental Material}
	
	\subsection{Halperin-Lubensky-Ma  for the dual easy-plane CP$^1$ model}
	
	The Halperin-Lubensky-Ma (HLM) mean-field theory \cite{HLM_PhysRevLett.32.292} is actually a calculation where mean-field theory is applied to an effective Higgs theory action where the gauge fields were integrated out exactly, something that it is only possible in the case of an Abelian Higgs model. An instance of it already existed in 3+1 dimension \cite{Coleman-Weinberg_PhysRevD.7.1888}, corresponding to a mechanism of inducing spontaneous symmetry breaking by quantum fluctuations. This symmetry breaking mechanism typically implies a first-order phase transition. In 2+1 dimensions it generates a non-analytical term in the effective potential, since assuming that the scalar field $\phi$ is uniform and integrating out the gauge field yields \cite{HLM_PhysRevLett.32.292}, 
	\begin{equation}
		{\rm Tr}\ln(-\partial^2+2e^2|\phi|^2)=\frac{2\Lambda e^2}{\pi^2}|\phi|^2-\frac{\sqrt{2}e^3}{\pi}|\phi|^3,
	\end{equation} 
	where $\Lambda$ is the UV cutoff, here assumed to be such that $\Lambda^2\gg 2e^2|\phi|^2$. The term $\sim |\phi|^3$ is the mentioned non-analytic term that causes the second-order phase transition from the Landau theory to turn into a first-order one. However, the presence of such a non-analytic term reveals that the essence of this problem is non-perturbative. The Dasgupta-Halperin duality \cite{Dasgupta-Halperin_PhysRevLett.47.1556} posits that in the strong coupling regime one actually finds a second-order phase transition. In order to see this, let us recall the continuum version of the dual model \cite{Kiometzis_PhysRevLett.73.1975}, 
	\begin{eqnarray}
		\mathcal{L}_{\rm dual}&=&\frac{1}{2}(\epsilon_{\mu\nu\lambda}\partial_\nu b_\lambda)
		^2+\frac{M^2}{2}b_\mu^2+|(\partial_\mu-iM\widetilde{e}b_\mu)\widetilde{\phi}|^2
		\nonumber\\
		&+&m^2|\widetilde{\phi}|^2+\frac{u}{2}|\widetilde{\phi}|^4, 
	\end{eqnarray}
	where  the scalar field $\widetilde{\phi}$ is dual to the original Higgs field $\phi$ and $\widetilde{e}=2\pi/e$ is the dual gauge coupling.
	This dual Lagrangian features a massive vector field $b_\mu$. Upon integrating out $b_\mu$, a term $\sim-(M^2+2\widetilde{e}^2|\widetilde{\phi}|^2)^{3/2}$ is generated. The latter leads to an analytic Landau expansion in $|\widetilde{\phi}|^2$. Due to the mass $M$, the interaction between vortex loops is screened and circumvents the first-order transition scenario from the HLM mean-field theory.
	
	The situation described above changes considerably for the $U(1)\times U(1)$ Abelian Higgs model. From the dual lattice action in Eq. (5) of the main text, one can infer a continuous field theory with the following Lagrangian, 
	\begin{eqnarray}
		\label{Eq:L-dual}
		\widetilde{\mathcal{L}}&=&\sum_{I=1,2}\left[
		\frac{1}{2}(\epsilon_{\mu\nu\lambda}\partial_\nu b_{I\lambda})^2
		+|(\partial_\mu-i\widetilde{e}~b_{I\mu})\phi_I|^2\right]
		\nonumber\\
		&+&\frac{M^2}{2}(b_{1\mu}+
		b_{2\mu})^2
		+m^2(|\phi_1|^2+|\phi_2|^2)
		\nonumber\\
		&+&\frac{u}{2}(|\phi_1|^4+|\phi_2|^4)
		+v|\phi_1|^2|\phi_2|^2,
	\end{eqnarray}
	where for the two-component case we have introduced a new dual bare gauge coupling $\widetilde{e}=\sqrt{2\pi/g}$ and 
	$M^2=e^2/g$. The theory  features two complex scalar fields $\phi_1$ and $\phi_2$ and two gauge fields, $b_{1\mu}$ and $b_{2\mu}$, along with a term $M^2(b_{1\mu}+b_{2\mu})^2/2$. In fact, integrating out both $b_{1\mu}$ and $b_{2\mu}$ yields an effective potential,  
	\begin{eqnarray}
		U_{\rm eff}(\phi_1,\phi_2)&=&\frac{\Lambda\widetilde{e}^2M^2}{3\pi^2}(|\phi_1|^2+|\phi_2|^2)
		\nonumber\\
		&-&\frac{M^3}{2\pi}\sum_{\sigma=\pm}\left[M_\sigma^2(\phi_1,\phi_2)\right]^{3/2}+\dots,
	\end{eqnarray}
	where,
	\begin{eqnarray}
		M_\sigma^2(\phi_1,\phi_2)&=&1+\widetilde{e}^2(|\phi_1|^2+|\phi_2|^2)
		\nonumber\\
		&\pm&\sqrt{1+\widetilde{e}^4(|\phi_1|^2-|\phi_2|^2)^2}.
	\end{eqnarray}
	Hence, up to a constant term, attempting to perform a Landau expansion gives us, 
	\begin{eqnarray}
		U_{\rm eff}(\phi_1,\phi_2)&=&\frac{M^2\widetilde{e}^2}{\pi}\left(\frac{\Lambda}{3\pi}-\frac{3M}{2\sqrt{2}}\right)(|\phi_1|^2+|\phi_2|^2)
		\nonumber\\
		&-&\frac{3M^3\widetilde{e}^4}{16\pi\sqrt{2}}\left[5(|\phi_1|^4+|\phi_2|^4)-6|\phi_1|^2|\phi_2|^2\right]
		\nonumber\\
		&-&\frac{M^3\widetilde{e}^3}{2\pi}(|\phi_1|^2+|\phi_2|^2)^{3/2}+\dots,
	\end{eqnarray}
	which also yields the non-analytic term characteristic of the first-order phase transition in the HLM mean-field theory. Thus, the easy-plane model features a non-analytic term both in the original and in the dual models. This result reflects the self-duality of the model. The reason why this happens can be easily understood by diagonalizing the gauge field matrix via the fields $b_{\pm\mu}=b_{1\mu}\pm b_{2\mu}$. Only $b_{+\mu}$ is gapped and contributes to screening of vortex loops, while $b_{-\mu}$ is gapless, leading to a HLM mean-field behavior like the one obtained from the original model by integrating out $a_\mu$.   There is henceforth a self-duality of the weak first-order transition described by the Halperin-Lubensky-Ma mechanism \cite{HLM_PhysRevLett.32.292}.      
	
	%{\color{blue} There is an interesting aspect regarding self-duality and first-order phase transitions in connection to the dual field theory of Eq. (\ref{Eq:L-dual}) that is worth discussing. A well known mean-field theory for the Abelian Higgs model due to Halperin {\it et al.} \cite{HLM_PhysRevLett.32.292} already suggests a weak first-order phase transition, since integrating out the gauge field for a uniform profile of the Higgs field $\psi$ generates a non-analytic contribution $\sim-|\psi|^3$ in the effective potential. On the other hand, a similar analysis in the dual field theory does not reproduce this behavior, since the gauge field is gapped \cite{Kiometzis_PhysRevLett.73.1975}. The same is not true in our case, since our dual field theory features two gauge fields and a term of the form $\sim(b_{1\mu}+b_{2\mu})^2$. Upon an appropriate rotation this leads to a gapped gauge mode and a gapless one. Thus, the weak first-order regime arising in the original model is reproduced in the dual field theory when both $b_{1\mu}$ and $b_{2\mu}$ are integrated out. There is henceforth a self-duality of the weak first-order transition described by the Halperin-Lubensky-Ma mechanism \cite{HLM_PhysRevLett.32.292}.     }
	
	Going one step further and accounting for the scalar field fluctuations, at one-loop order the RG equations for dimensionless couplings $\hat{u}$ and $\hat{v}$ yield
	\begin{eqnarray}
		\mu \frac{d \hat{u}}{d \mu}&=&-(4-d) \hat{u}+2\left[(N+4) \hat{u}^{2}+N \hat{v}^{2}+2(d-1) f^{2}\right]\nonumber\\
		\mu \frac{d \hat{v}}{d \mu}&=&-(4-d) \hat{v}+4 \hat{v}^{2}+4(N+1) \hat{u} \hat{v},
	\end{eqnarray}
	where we used a notation $ f = \widetilde{e}^2\mu^{d-4}$ and $\mu$ is a renormalization scale. The dimensionless  gauge coupling $f$ has the following $\beta$ function,
	\begin{equation}
		\mu \frac{d f}{d \mu}=-(4-d)f+\frac{f^{2}}{24 \pi}.
	\end{equation}
	In our case of  $N=1$ and $d=3$, there are no real solutions for this system of equations if $f$ is nonzero. A runaway flow is obtained and no second-order phase transition occurs, similarly to Ref. \cite{HLM_PhysRevLett.32.292}. 
	
	%The gauge coupling $h^2$ gets renormalized via the vacuum polarization (Fig. \ref{Fig:poldiag}). At  one-loop order, the renormalized coupling is given by
	%\begin{equation}
	%h_{R}^{2}=h^2-\frac{h^{4}}{3(4 \pi)^{d / 2}} %\left.\Gamma\left(2-\frac{d}{2}\right) m_R^{d-4}\right| %_{d=3}=h^2-\frac{h^{4}}{24 \pi m_R}.
	%\end{equation}
	
	%\begin{figure}[h]
	%	\includegraphics[width=4cm]{graph4p-1-1.pdf}
	%	\caption{Vacuum polarization diagram. The wiggles represent the gauge field. }
	%	\label{Fig:poldiag}
	%\end{figure}
	
	%Defining the dimensionless coupling $\hat{h}^2 = 	h_{R}^{2}/m_R$, one obtains an RG equation,
	%\begin{equation}
	%	m_R \frac{d \hat{h}^{2}}{d m_R}=-\hat{h}^2+\frac{\hat{h}^{4}}{24 \pi}.
	%\end{equation}

	\subsection{Renormalization group analysis of the dual model}	
	
	Here we will perform an RG analysis of the continuous Lagrangian of the dual model presented in Eq. (11) of the main text,
	\begin{eqnarray}
		\label{Eq:L-semidual}
		\widetilde{\mathcal{L}}''&=&
		|(\partial_\mu-iea_{\mu})\phi_1|^2+m^2|\phi_1|^2+ \frac{u_1}{2}|\phi_1|^4 \nonumber\\
		&+&|(\partial_\mu-i\widetilde{e}b_{\mu})\phi_2|^2+m^2|\phi_2|^2+ \frac{u_2}{2}|\phi_2|^4 
		\nonumber\\
		&+&\frac{1}{2}\left(\varepsilon_{\mu \nu \lambda} \partial_{\nu} a_{ \lambda}\right)^{2}+ \frac{1}{2}\left(\varepsilon_{\mu \nu \lambda} \partial_{\nu} b_{ \lambda}\right)^{2}\nonumber\\&+&\frac{i e}{\sqrt{g}} a_{\mu} \epsilon_{\mu\nu\lambda}\partial_\nu b_{\lambda},
	\end{eqnarray}
	where the charge $\widetilde{e}$ is defined in terms of the original coupling as $\widetilde{e} = \frac{2 \pi}{\sqrt{g}}$.
	
\begin{figure}[htbp]
		\centering
		\includegraphics[width=3cm]{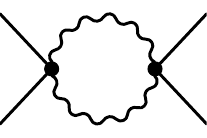}
		\caption{Gauge field bubble diagram contributing to the coupling $u_r$. 
			External lines represent either $\phi_1$ or $\phi_2$. The wiggle represents 
			either $b_{1\mu}$ or $b_{2\mu}$.}
		\label{Fig:photon4p-loop}
	\end{figure}

\begin{figure}[htpb]
		\includegraphics[width=4cm]{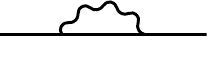}
		\caption{Scalar field self-energy}
		\label{Fig:self-energy}
	\end{figure}

	Integrating out the gauge fields in the Lagrangian \eqref{Eq:L-semidual}, we calculate a matrix gauge field propagator,
	\begin{equation}
		\mathbb{D}_{\mu\nu}(p)=\left[
		\begin{array}{cc}
			\frac{1}{p^2+M^2}\left(\delta_{\mu \nu}-\frac{p_\mu p_\nu}{p^2}\right) & M\frac{\epsilon_{\mu \nu \lambda}p_\lambda}{p^2(p^2+M^2)}\\
			\noalign{\medskip}
			M\frac{\epsilon_{\mu \nu \lambda}p_\lambda}{p^2(p^2+M^2)} & 	\frac{1}{p^2+M^2}\left(\delta_{\mu \nu}-\frac{p_\mu p_\nu}{p^2}\right)
		\end{array}
		\right],\\
	\end{equation}
	where $M^2 = e^2/g$ and we used the Landau gauge. The diagonal element of the matrix propagator allows us to calculate the contribution from the bubble diagram (Fig. \ref{Fig:photon4p-loop}) and a self-energy (Fig. \ref{Fig:self-energy}) that enters the wave function renormalization. 
	
	As the gauge fields have different charges, the gauge field bubble diagrams provide contributions evaluated by the integral of the following form,
	\begin{equation}
		2 s_{1} h^{4} \int \frac{1}{\left(p^{2}+M^2\right)^{2}} = \frac{s_1 h^4}{4 \pi M},
	\end{equation}
	where $h^2$ plays the role of $e^2$ or $\widetilde{e}^2$ and $s_1$ is a symmetry factor of the diagram that is found to be equal to 2. Therefore, the gauge field $a_\mu$ bubble diagram results in $e^3 \sqrt{g}/(2 \pi)$, while for $b_\mu$ the contribution is equal to $8\pi^3/(e g^{3/2})$. 
	
	The self-energy diagram (Fig. \ref{Fig:self-energy}) contributes to the wave function renormalization through the expansion up to a $p^2$ term. The diagram corresponds to the integral
	\begin{equation}
		-4 h^{2} \int_{k} \frac{p_{\mu} p_{\nu}}{\left[(p-k)^{2}+m^{2}\right]\left(k^{2}+M^{2}\right)}\left(\delta_{\mu}-\frac{k_{\mu} k_{\nu}}{k^{2}}\right),
	\end{equation}
	which is well known and the result of the integration can be found in the Appendix A.2 of \cite{Nogueira2019}. The wave function renormalizations  corresponding to the two gauge fields have the same form,
	\begin{equation}
		Z=1+\frac{2 h^{2}}{3 \pi} \frac{1}{m_R+M},
	\end{equation}
	where to evaluate the contributions for $a_\mu$ or $b_\mu$, one has to substitute the coupling $h$ with $e$ or $\widetilde{e}$, respectively.

	To express the renormalized couplings $u_{1R}$ and  $u_{2R}$, one needs to calculate the so called fish diagram. In the case of both couplings the contribution is equal to $(s_2 u_I^2)/(8 \pi m_R)$, where $I=1,2$ and $s_2 = 5$ is a symmetry factor of the diagram.   
	\begin{widetext}

		Eventually, one obtains a renormalized coupling $u_{1R}$ and can define the dimensionless coupling $\hat{u}_1$,
		\begin{eqnarray}
			\hat{u}_1&=&\frac{u_{1R}}{m_R}=Z_1^{2}\left(\frac{u_1}{m_R}-\frac{5 u_1^{2}}{8 \pi m_R^{2}}-\frac{e^3 \sqrt{g}}{2 \pi{m_R}}\right)\\&\approx&\frac{u_{1}}{m_{R}}\left(1+\frac{4 e^{2}}{3 \pi} \frac{1}{m_{R}+\frac{e}{\sqrt{g}}}\right)-\frac{5 u_{1}^{2}}{8 \pi m_{R}^{2}}-\frac{e^{3} \sqrt{g}}{2 \pi m_{R}}.\nonumber
		\end{eqnarray}
		
		Calculating the $\beta$ function for $\hat{u}_1$, one obtains
		\begin{eqnarray}
			m_R \frac{d \hat{u}_1}{d m_R}&=&-\frac{u_{1}}{m_{R}}\left(1+\frac{4 e^{2}}{3 \pi} \frac{1}{m_{R}+M}\right)+\frac{5 u_{1}^{2}}{4 \pi m_{R}^{2}}+\frac{e^{3} \sqrt{g}}{2 \pi m_{R}}\nonumber \\&=& -\underbrace{\left[\frac{u_{1}}{m_{R}}\left(1+\frac{4 e^{2}}{3 \pi} \frac{1}{m_{R}+M}\right)-\frac{5 u_{1}^{2}}{8 \pi m_{R}^{2}}-\frac{e^{3} \sqrt{g}}{2 \pi m_{R}}\right]}_{=\hat{u}_1}+\frac{5 u_1^{2}}{8 \pi m_R^{2}}\nonumber \\&=& -\hat{u}_1  +\frac{5 u_1^{2}}{8 \pi m_R^{2}} \approx - \hat{u}_1  +\frac{5 \hat{u}_1^{2}}{8 \pi}.
		\end{eqnarray}
		In a similar fashion, we evaluate the $\beta$ function for $\hat{u}_2$,
		\begin{eqnarray}
			m_R \frac{d \hat{u}_2}{d m_R}&=&-\frac{u_{2}}{m_{R}}\left(1+\frac{16 \pi}{3 g} \frac{1}{m_{R}+M}\right)+\frac{10 u_{2}^{2}}{8 \pi m_{R}^{2}}+\frac{8 \pi^{3}}{e g^{3 / 2} m_{R}}\nonumber \\&=& -\underbrace{\left[\frac{u_{2}}{m_{R}}\left(1+\frac{16 \pi}{3 g} \frac{1}{m_{R}+M}\right)-\frac{5 u_{2}^{2}}{8 \pi m_{R}^{2}}-\frac{8 \pi^{3}}{e g^{3 / 2} m_{R}}\right]}_{=\hat{u}_2}+\frac{5 u_2^{2}}{8 \pi m_R^{2}}\nonumber \\&=& -\hat{u}_2  +\frac{5 u_2^{2}}{8 \pi m_R^{2}} \approx - \hat{u}_2  +\frac{5 \hat{u}_2^{2}}{8 \pi}.
		\end{eqnarray}
	\end{widetext}

	This way we obtain $\beta_{\hat{u}_I}$ which vanish at $\hat{u}_I = 0$ and $\hat{u}_I = (8\pi)/5$, where $I=1,2$. Note that this RG analysis employed the fixed dimension approach pioneered by Parisi \cite{Parisi}, which is at first sight less controlled, since there is no fixed point of $\mathcal{O}(\epsilon)$. However, this is not an actual concern, since the the perturbation series has to be resummed anyway at higher orders. Furthermore, we could in principle consider the $\epsilon$ expansion as well, and the same result would have followed.

	\subsection{Bosonization through flux attachment}
	To see how the existence of the critical point in the easy-plane CP$^1$ model has consequences for fermionic systems, we turn to the flux attachment technique to derive a fermionized version of the bosonic model considered in the main text. It was demonstrated that the frozen limit ($g\to0$) of the model with both $U(1)$ sectors dualized almost completely coincides with the partially dual model where $g$ is kept finite. At the critical point the Maxwell terms arising in the partially dual model become irrelevant and so both field theories have the same form,
	\begin{eqnarray}
		\mathcal{L}_b&=&\sum_{I=1,2}\left[|(\partial_\mu-ib_{I\mu})\phi_I|^2 + m^2 |\phi_I|^2 + \frac{u}{2}|\phi_I|^4 \right]\nonumber\\&+& \frac{i}{2\pi}\epsilon_{\mu\nu\lambda}b_{1\mu}\partial_\nu b_{2\lambda},
	\end{eqnarray}
	up to a sign of the BF term. The latter is, however, immaterial and does not play any role in the bosonization process we perform below.
	
	We start with the following well known conjecture  \cite{Karch_PhysRevX.6.031043, Senthil_PhysRevX.7.031051, SEIBERG2016395}, 
	\begin{equation}
		\label{Eq:Fermion+flux}
		Z_{\rm f Q E D+flux}[A]=Z_{\rm b Q E D}[A] e^{S_{CS}[A]},
	\end{equation}
	written in the imaginary time formalism. The flux attachment used in the conjecture has the form of,
	\begin{eqnarray}
		\label{Eq:ZfQED+flux}
		Z_{\rm fQED+flux}[A]&=&\int\mathcal{D}a_\mu 	Z_{\rm fQED}[a]e^{-\frac{1}{2}S_{CS}[a] -S_{BF}[a;A]},
		\nonumber\\
	\end{eqnarray}
	where
	\begin{eqnarray}
		\label{Eq:ZfQED}
		Z_{\rm fQED}[A]&=&\int\mathcal{D}\bar{\psi}\mathcal{D}\psi e^{-S_{\rm fQED}[A]},
		\nonumber\\
		S_{\rm fQED}[A]&=&\int d^3x\bar{\psi}(\slashchar{\partial}-i\slashchar{A})\psi,	\nonumber\\	
		S_{CS}[A]&=&\frac{i}{4 \pi} \int d^3x\epsilon_{\mu\nu\lambda}A_{\mu}\partial_\nu A_{\lambda},
	\end{eqnarray}
	with the latter being a topological Abelian Chern-Simons (CS) action.
	
	The duality we are interested in involves two bosonic fields. To account for this, we use the conjecture \eqref{Eq:Fermion+flux} twice,
	\begin{eqnarray}
		\label{Eq:2comp}
		&&Z_{\rm f Q E D+flux}[A_1] Z_{\rm f Q E D+flux}[A_2]  e^{-S_{CS}[A_1]-S_{CS}[A_2]}\nonumber\\&=&Z_{\rm b Q E D}[A_1] Z_{\rm b Q E D}[A_2].
	\end{eqnarray}
	The technique we use here has recently been applied to find a fermionic dual of the topological version of the easy-plane CP$^1$ model \cite{Shyta_PhysRevLett.127.045701, Shyta_PhysRevD.105.065019}. The results from the flux attachment were shown to agree with the exact duality transformations in the spirit of Refs. \cite{Peskin1978, Dasgupta-Halperin_PhysRevLett.47.1556}. 
	
	Since the bosonic theory is not supposed to contain any CS terms, the latter now appear on the fermionic side of the conjecture. 
	
	We multiply both sides of the expression above by $\exp{(- S_{BF}[A_1; A_2])}$ and promote the background fields $A_1$ and $A_2$ to be dynamical  $b_1$ and $b_2$. This promotion requires introducing two new background fields, which we will denote as  $C_1$ and $C_2$. Then, the expression \eqref{Eq:2comp} takes the form,
	\begin{widetext}
		\begin{eqnarray}
			\label{eq:promoted2A}
			&& \int\mathcal{D}b_{1\mu}\mathcal{D}b_{2\mu}Z_{\rm fQED + flux}[b_1]Z_{\rm fQED+flux}[b_2]e^{-S_{CS}[b_1+b_2]+S_{BF}[b_1;C_1]+S_{BF}[b_2;C_2]} \nonumber\\&=&
			\int\mathcal{D}b_{1\mu}\mathcal{D}b_{2\mu}Z_{\rm bQED}[b_1]Z_{\rm bQED}[b_2]e^{-S_{BF}[b_1; b_2] + S_{BF}[b_1;C_1]+S_{BF}[b_2;C_2]}.
		\end{eqnarray}
		Using the definition of the fermionic flux attachment, we integrate out the dynamic gauge fields $b_1$ and $b_2$ on the fermionic side of the duality in Eq. \eqref{eq:promoted2A}. We arrive at the expression,
		\begin{eqnarray}
			\label{eq:integrated2b}
			&& \int\mathcal{D}a_{\mu}Z_{\rm fQED }[a]Z_{\rm fQED}[a+C_2-C_1]e^{-\frac{1}{2}S_{BF}[a; C_1+C_2]-\frac{1}{2}S_{CS}[C_2-C_1]+S_{CS}[C_1]}\nonumber\\&=&
			\int\mathcal{D}b_{1\mu}\mathcal{D}b_{2\mu}Z_{\rm bQED}[b_1]Z_{\rm bQED}[b_2]e^{-S_{BF}[b_1; b_2] + S_{BF}[b_1;C_1]+S_{BF}[b_2;C_2]}.
		\end{eqnarray}
		To make the fermionic side of duality more symmetrical, we perform a shift $a\to a+ (C_1-C_2)/2$,
		\begin{eqnarray}
			\label{eq:integrated2bsym}
			&& \int\mathcal{D}a_{\mu}Z_{\rm fQED }\left[a-(C_2-C_1)/2\right]Z_{\rm fQED}\left[a+(C_2-C_1)/2\right]e^{-\frac{1}{2}S_{BF}[a; C_1+C_2]-S_{CS}[C_2]+S_{CS}[C_1] +\frac{1}{2}S_{BF}[C_1;C_2]}\nonumber\\&=&
			\int\mathcal{D}b_{1\mu}\mathcal{D}b_{2\mu}Z_{\rm bQED}[b_1]Z_{\rm bQED}[b_2]e^{-S_{BF}[b_1; b_2] + S_{BF}[b_1;C_1]+S_{BF}[b_2;C_2]}.
		\end{eqnarray}
		And so, the bosonization duality relates a bosonic theory with two interacting dynamical gauge fields to the theory of two massless Dirac fermions coupled via the same gauge field,  
		\begin{eqnarray}
			\label{Eq:MainBosonization}
			\mathcal{L}_b&=&\sum_{I=1,2}\left[|(\partial_\mu-ib_{I\mu})\phi_I|^2 + m^2 |\phi_I|^2 + \frac{u}{2}|\phi_I|^4 \right]+\frac{i}{2\pi}\epsilon_{\mu\nu\lambda}b_{1\mu}\partial_\nu b_{2\lambda}
			\nonumber\\
			&&\hspace{2cm} \big\Updownarrow
			\nonumber\\
			\mathcal{L}_f&=&\sum_{I=1,2} \bar{\psi}_I(\slashchar{\partial}-i\slashchar{a})\psi_I,
		\end{eqnarray}
		where we put the background fields to zero. 
	\end{widetext}

\end{document}